\documentclass[physrev,reprint,superscriptaddress,bibnotes]{revtex4-2}

\usepackage[T1]{fontenc}
\usepackage[american]{babel}

\usepackage{amsmath,amsthm,amssymb,graphicx,bm,microtype}
\usepackage[dvipsnames]{xcolor}
\usepackage[colorlinks,allcolors=blue!50!black]{hyperref}
\usepackage[all]{hypcap}
\usepackage{cleveref}

\usepackage{orcidlink}
\newcommand{\orcid}[1]{\,\orcidlink{#1}}

\usepackage{braket}
\usepackage[normalem]{ulem}

\newcommand{\avg}[1]{\langle #1 \rangle}
\newcommand{\cop}[0]{\hat{c}}
\newcommand{\fop}[0]{\hat{f}}

\newcommand{\nop}[1]{\hat{n}_{#1}}

\def\B{\ensuremath\mathcal{B}}

\let\Re\relax \DeclareMathOperator{\Re}{Re}
\let\Im\relax \DeclareMathOperator{\Im}{Im}

\newtheorem{lemma}{Lemma}

\newcommand{\UQ}{School of Mathematics and Physics, The University of Queensland, 4072, Australia}
\newcommand{\USYD}{School of Chemistry, University of Sydney, NSW 2006, Australia}

\begin{document}

\title{Accelerating dynamical mean-field theory convergence by preconditioning with computationally cheaper quantum embedding methods}

\author{E. M. Makaresz\orcid{0009-0001-1243-7109}}
\affiliation{\UQ}
\author{O. Gingras\orcid{0000-0003-3970-6273}}
\affiliation{Center for Computational Quantum Physics, Flatiron Institute, 162 Fifth Avenue, New York, New York 10010, USA}
\affiliation{Université Paris-Saclay, CNRS, CEA, Institut de physique théorique, 91191, Gif-sur-Yvette, France}
\author{Tsung-Han Lee}
\affiliation{Department of Physics, National Chung Cheng University, Chiayi 62102, Taiwan}
\author{Nicola Lanat\`a\orcid{0000-0003-0003-4908}}
\affiliation{School of Physics and Astronomy, Rochester Institute of Technology, Rochester, New York 14623, USA}
\affiliation{Center for Computational Quantum Physics, Flatiron Institute, 162 Fifth Avenue, New York, New York 10010, USA}
\author{B. J. Powell\orcid{0000-0002-5161-1317}}
\affiliation{\UQ}
\author{Henry L. Nourse\orcid{0000-0003-3610-1105}}
\affiliation{\USYD}

\date{\today}

\begin{abstract}
    Dynamical mean-field theory (DMFT) is a cornerstone technique for studying strongly correlated electronic systems. However, each DMFT step is computationally demanding, and many iterations can be required to achieve convergence. Here, we accelerate the convergence of DMFT by initializing its self-consistent cycle with solutions from computationally cheaper and more approximate methods. We compare the initialization with the non-interacting solution to a range of quantum embedding compatible approaches: Hartree-Fock, the Hubbard-I approximation, rotationally invariant slave bosons (RISB), and its ghost extension (g-RISB). We find that these initializations can reduce the number of DMFT iterations by up to an order of magnitude, with g-RISB providing the most effective and reliable benefits. In most regimes, initializing with g-RISB and performing a single DMFT iteration suffices to recover the full dynamical structure. The improvement in convergence is controlled by the initial solution’s accuracy in the low-energy part of the self-energy, on the scale of the non-interacting bandwidth. This strategy is especially effective at the Mott insulator-metal transition, where an initialization from the non-interacting limit can lead to a breakdown of DMFT due to the sign problem. Our results establish the usage of accurate yet cheaper quantum embedding methods as a powerful means to substantially reduce the computational cost of DMFT, particularly in regimes where convergence is slow or prone to failure.
\end{abstract}

\maketitle

\section{Introduction}

Strong electronic correlations lie at the foundation of quantum materials~\cite{Quintanilla_2009,LeeNW}, giving rise to a wide range of emergent phenomena~\cite{hunds_coupling_Georges_2013,hshim2007modeling,dzero2016topological,kotliar_etal_dmft_review_2006, dft+dmft_applications_2019}. These correlated states are central to the development of a range of next-generation quantum technologies~\cite{Tokura_quantum_materials_review_2017,Milloch,Wang_resistive_switching_review_2020}, with important applications including quantum sensing~\cite{Milloch, Tokura_quantum_materials_review_2017}, energy-efficient memory and logic devices~\cite{Milloch, spin_colossal_magnetoresistance_Qiu_2018, Tokura_quantum_materials_review_2017, Wang_resistive_switching_review_2020}, and spintronic applications~\cite{Milloch, spin_colossal_magnetoresistance_Qiu_2018, Tokura_quantum_materials_review_2017}. They also hold promise for emerging technologies aimed at decarbonization, including improved energy storage and carbon capture and conversion~\cite{porous_MOF_review_2016, Tokura_quantum_materials_review_2017, Furukawa_MOF_review_2013, Deanna-review}. Despite this broad technological potential, their theoretical description remains a major challenge: accurately capturing the interplay of strong many-body interactions requires computational resources that scale exponentially with system size, severely limiting predictive simulations of realistic materials.

Dynamical mean-field theory (DMFT)~\cite{InfiniteD,kotliar_etal_dmft_review_2006} has become a cornerstone for studying strongly correlated systems, making their realistic simulation feasible and achieving quantitative agreement with experiments on specific materials~\cite{kotliar_etal_dmft_review_2006, dft+dmft_applications_2019,transition-metal_oxide_design_2021,yin2011magnetism,fernandes2022iron}. By mapping the full interacting problem onto a self-consistent quantum impurity model, DMFT captures local dynamical correlations exactly while neglecting nonlocal ones, reducing the original many-body problem to a numerically solvable impurity embedded in a self-consistent bath. The impurity self-energy is then fed back into the interacting Green’s function of the system, and the procedure is iterated until self-consistent convergence is reached. 

However, the accuracy of DMFT carries substantial computational cost: although solving the impurity model is much simpler than the full interacting one, each DMFT iteration demands considerable resources when using numerically accurate impurity solvers~\cite{CTQMC_2011, NRG_review_2008, iNRG_2016, ED_imp_solver_2017}. Convergence can become prohibitively slow in large or spatially inhomogeneous systems~\cite{decoupled_impurities_1999, Jaime_decoupled_dmft, rr-DMFT,kotliar_etal_dmft_review_2006, dft+dmft_applications_2019} and in regimes where the fermionic sign problem limits numerical accuracy~\cite{DMFT_book,CTQMC_2011}, constraining its routine use in realistic materials simulations and low-temperature regimes.

In this work, we propose a simple and general strategy to make DMFT calculations more efficient. Instead of initializing DMFT from the uncorrelated state, we start the self-consistency from the self-energy obtained using a computationally cheaper correlated method. Using the single-orbital Hubbard model on the square lattice as a case study, we show that initialization with a correlated method can improve convergence by an order of magnitude. We note that, when the initializer captures the essential low-energy properties of the system, only a single DMFT iteration can accurately reproduce the converged solution. Among the methods benchmarked, initializing with the ghost-rotationally invariant slave-boson method~\cite{original_g-risb, g-risb_benchmark_1, g-risb_benchmark_2,gGA1_2017,gGA2_2021} requires the fewest number of DMFT iterations to reach convergence across all parameters.

\section{Improving the initialization of DMFT}\label{sect:dmft}

DMFT is an iterative quantum embedding framework which is exact in zero and infinite dimensions and therefore exactly captures all local electronic correlations. It works by iteratively mapping the full lattice to an effective impurity model hybridized to a non-interacting bath, solving this simpler many-body problem and embedding its solution back into the original lattice. Its solution is found when the self-consistency condition is satisfied, that is when the local Green's function of the full interacting system, $G^{\text{loc}}$, equals the Green's function of the effective impurity problem, $G^{\text{imp}}$~\cite{InfiniteD,DMFT_review_1996}.

In a calculation with a lattice dispersion $\varepsilon_{\bm{k}}$ and a fixed on-site occupation, one has to work in the grand canonical ensemble and needs to find the right chemical potential $\mu$. In that case, the DMFT self-consistency loop is initialized with a trial self-energy $\Sigma$ and trial chemical potential $\mu$. Using this, the local interacting Green's function can be computed as
\begin{equation} \label{eq:g-loc-dmft}
    G^{\text{loc}}(z) = \frac{1}{N}\sum_{\bm{k}} \frac{1}{z + \mu - \varepsilon_{\bm{k}} - \Sigma(z)},
\end{equation}
where $N$ is the number of $\bm{k}$-points in the first Brillouin zone, and $z$ is a complex frequency which is typically either a real frequency plus a small imaginary broadening $\omega + i\eta$, or a purely imaginary Matsubara frequency $i\omega_n$. 

The next step is to construct the effective impurity problem. This is done by considering the local interacting Hamiltonian $\hat{H}^{\mathrm{int}}$, and by defining the Weiss field, which corresponds to the non-interacting impurity Green’s function, $G^{\text{imp}}_0$, obtained via Dyson’s equation~\cite{DMFT_book}
as 
\begin{equation}\label{eq:weiss-dmft}
    G_0^{\text{imp}}(z) = \frac{1}{[G^{\text{loc}}(z)]^{-1} + \Sigma(z)}.
\end{equation}
This impurity problem is then solved using a so-called impurity solver to obtain the interacting impurity Green's function $G^{\mathrm{imp}}(z)$. From this, a new self-energy $\Sigma(z)$ can be extracted and embedded back into the original lattice.
There, a new chemical potential and an updated local interacting Green's function can be calculated, starting the next iteration of the self-consistency cycle. This cycle is repeated until the chemical potential $\mu$ and the Weiss field $G_0^{\mathrm{imp}}$ are converged.

Only the low-frequency structure of $\Sigma(z)$ controls the proximity to the DMFT fixed point, since the Weiss field is insensitive to the high-frequency part of the self-energy. Indeed, this can be seen by considering the bound $|\varepsilon_{\bm{k}} - \varepsilon_0| \leq W$, where 
\begin{equation}
    \label{eq:epsilon_0}
    \varepsilon_0 = \frac{1}{N} \sum_{\bm{k}} \varepsilon_{\bm{k}}
\end{equation}
and $W$ is the non-interacting bandwidth, and approximating the lattice dispersion at high frequencies $|z| \gg W$ by keeping only its local contribution, $z - \varepsilon_{\bm{k}} \approx z - \varepsilon_0$. In this high-frequency regime, the Weiss field becomes independent of the self-energy:
\begin{align}\label{eq:sigma_cancellation}
    \nonumber G_0^{\mathrm{imp}}(z) &\approx \left[ \left( \frac{1}{N} \sum_{\bm{k}} \frac{1}{z + \mu - \varepsilon_0 - \Sigma(z)} \right)^{-1} + \Sigma(z) \right]^{-1}
    \\
    &= \frac{1}{z + \mu - \varepsilon_0} \quad \textrm{for } |z| \gg W.
\end{align}
Thus, an effective initial self-energy must reproduce the low-energy correlation effects on the order of the bandwidth $W$, while its performance is largely insensitive to its accuracy at higher frequencies. 

Based on this observation, we propose the following approach to accelerate and reduce the computational cost of self-consistent DMFT calculations:
\begin{enumerate}
    \item Solve the target system using a computationally cheaper method and extract the self-energy $\Sigma$ and chemical potential $\mu$.
    \item Initialize the DMFT self-consistency loop with these quantities.
    \item Using a numerically accurate impurity solver, iterate to full convergence or, if the solution is close to the DMFT converged solution, perform only a single DMFT iteration.
\end{enumerate}
This procedure bypasses the bulk part of the self-consistent procedure by providing a better starting point and therefore achieves the same results at significantly lower cost.

\section{Methods used for initialization}\label{sect:methods}

Various other schemes than DMFT have been developed to study strongly correlated systems at a lower cost, including simplified impurity solvers for DMFT~\cite{HubbardI_original,cluster_NCA,multiorbital_IPT_2016}, variational methods~\cite{gutzwiller1963effect,gutzwiller1964effect,metzner1989correlated,bunemann1998multiband,nicola2012efficient,gGA1_2017,gGA2_2021,knizia2012density,knizia2013density}, and auxiliary-particle formulations~\cite{KRSB_original_1986,original_risb_2007,original_g-risb,slave_rotor_review_2004,medici2005orbital-selective,ruegg2010z2-slave-spin}. Among these, ghost rotationally invariant slave bosons (g-RISB)~\cite{original_g-risb,g-risb_benchmark_1,g-risb_benchmark_2}---or equivalently the ghost Gutzwiller approximation (gGA)~\cite{gGA1_2017,gGA2_2021}---stands out as particularly promising. Similarly to DMFT, g-RISB is formally a quantum embedding method~\cite{original_g-risb}. Although it does not achieve the full dynamical accuracy of DMFT, it has been shown to capture some incoherent effects and reproduce key low-energy properties with high, systematically improvable accuracy~\cite{original_risb_2007,g-risb_benchmark_1,g-risb_benchmark_2,risb_hund_study_2018}.

We consider three methods to initialize DMFT other than g-RISB, each with simple open-source implementations: Hartree--Fock (HF)~\cite{Hartree_Fock_original}, Hubbard-I (HI)~\cite{HubbardI_original} and rotationally invariant slave bosons (RISB)~\cite{original_risb_2007,risb_impurity_mapping_2015,KRSB_original_1986}. We benchmark these against the standard practice of initializing DMFT with the non-interacting (NI) solution. We have chosen these methods because each is useful in its characteristic regime of electronic correlations---HF captures weakly correlated behavior, HI describes the strongly localized limit, RISB effectively treats correlated metals, and g-RISB extends this accuracy across the metal–insulator transition, remaining reliable even in the correlated insulating regime.

\subsection{Paramagnetic Hubbard model}
\label{sect:methods-model}

In this work, we focus on the \emph{paramagnetic} solutions of the single-orbital Hubbard model to clearly isolate the competition between kinetic energy and local interactions that drive the Mott insulator-metal transition, but our approach applies equally to multi-orbital Hamiltonians and realistic first-principles calculations. 

The Hubbard model is defined by the Hamiltonian
\begin{equation}\label{eq:hubbard-model}
\hat{H} = \sum_{\bm{k}\sigma}(\varepsilon_{\bm{k}} - \mu) \cop^{\dagger}_{\bm{k}\sigma} \cop_{\bm{k}\sigma} + U\sum_{i} \hat{n}_{i\uparrow} \hat{n}_{i\downarrow},
\end{equation}
where $\cop^{\dagger}_{\bm{k}\sigma}$ creates and $\cop_{\bm{k}\sigma}$ annihilates an electron with crystal momentum $\bm{k}$ and spin $\sigma \in \{\uparrow,\downarrow\}$, $\cop^{\dagger}_{i\sigma}$ and $\cop_{i\sigma}$ do likewise for site $i$, $\hat{n}_{i\sigma}= \cop^{\dagger}_{i\sigma} c_{i\sigma}$ is the number operator, $\mu$ is the chemical potential, and $U$ is the on-site Coulomb interaction. 

We consider a square lattice with nearest and second nearest neighbors, leading to the following dispersion:
\begin{multline}
    \varepsilon_{\bm{k}} = -2t\big( \cos(k_x) + \cos(k_y) \big) \\
    - 2t'\big( \cos(k_x + k_y) + \cos(k_x - k_y) \big),
\end{multline}
where $t$ and $t'$ are the nearest- and next-nearest-neighbor hoppings. We set $t'/t = 0.15$ to avoid the special cases of particle–hole symmetry and perfect nesting, giving a non-interacting bandwidth $W = 8t$ and a critical interaction $U_c/W \approx 1.3$ for the Mott insulator-metal transition at half filling.

\subsection{Hartree-Fock}
\label{sect:methods-HF}

The Hartree-Fock (HF) approximation~\cite{Hartree_Fock_original} is generally considered the simplest framework for treating interactions in many-body systems.
It amounts to writing the operators of the interaction as a mean-value plus fluctuations, and truncating the terms where the fluctuations are of order two or higher.
More specifically, for the Hubbard-like interaction in Eq.~(\ref{eq:hubbard-model}), it amounts to writing $\hat{n}_{i\sigma} = \langle \hat{n}_{i\sigma} \rangle + (\hat{n}_{i\sigma} - \langle \hat{n}_{i\sigma} \rangle)$ and thus approximating the interaction as
\begin{align}
    \label{eq:HF_mean-field}
    \hat{n}_{i\uparrow} \hat{n}_{i\downarrow} \approx \hat{n}_{i\uparrow} \langle \hat{n}_{i\downarrow} \rangle + \hat{n}_{i\downarrow} \langle \hat{n}_{i\uparrow} \rangle - \langle \hat{n}_{i\uparrow}\rangle \langle \hat{n}_{i\downarrow}\rangle.
\end{align}

As a result, HF replaces the quartic interaction terms with bilinear ones, reducing the Hamiltonian to a quadratic form. The resulting effective single-particle Hamiltonian can then easily be diagonalized, with the averages updated self-consistently until convergence is reached.

In paramagnetic HF, the interaction generates a static, frequency-independent self-energy that reflects only the mean-field potential from the average electronic occupations. For the paramagnetic Hubbard model, this reduces to a local self-energy,
\begin{equation} \label{eq:self-energy-HF}
\Sigma = U \frac{n}{2}, 
\end{equation}
known as the Hartree shift, where $n / 2 = \avg{\hat{n}_{i\uparrow}} = \avg{\hat{n}_{i\downarrow}}$. This simply shifts all the single-particle energies without altering their dispersion and acts as a change to the chemical potential.
The last term on the right hand-side of Eq.~(\ref{eq:HF_mean-field}) only shifts the total energy and has no other physical effect.

In contrast to DMFT, HF neglects dynamical correlations entirely, as its self-energy lacks frequency dependence. As a result, it omits both the local and temporal fluctuations that are central to strongly correlated regimes, failing to capture both the correlation-driven insulating behavior of paramagnets and the renormalization effects essential for describing correlated metals~\cite{DMFT_book}. Despite these limitations, HF remains a valuable baseline method as it is computationally inexpensive and highlights the correlated physics that more advanced approaches capture.

\subsection{Hubbard-I}
\label{sect:methods-HI}

The Hubbard-I (HI) approximation~\cite{HubbardI_original} can be employed as a simple impurity solver within DMFT: rather than solving the full impurity problem with a dynamical bath, HI neglects the impurity-bath hybridization entirely and treats the impurity as an isolated atom. This makes HI far less computationally demanding compared to numerically exact DMFT, since only the atomic Green’s function corresponding to $\hat{H}^{\mathrm{int}}$ must be solved.

The self-energy in HI arises entirely from
the atomic problem, capturing only transitions between the eigenstates of the isolated impurity. For the paramagnetic Hubbard model, it takes the form~\cite{DMFT_book}
\begin{equation} \label{eq:self-energy-HI}
\Sigma(z) = U \frac{n}{2} + \frac{U^2 \tfrac{n}{2} (1 - \tfrac{n}{2})}{z + \mu - U(1-\tfrac{n}{2})}.
\end{equation}
The first term represents the static Hartree shift, while the second introduces a coherent pole at $z = U(1 - n /2) - \mu$, corresponding to charge excitations between atomic multiplets. This pole splits the electronic spectrum into lower and upper Hubbard bands whenever $U>0$, producing the characteristic spectrum of a Mott insulator.

This atomic form of the self-energy determines both the strength and the limitations of HI. Because it contains only transitions between localized multiplet states, the approximation captures the insulating limit where electrons are immobile, but it fails to reproduce the coherent quasiparticle excitations characteristic of correlated metals.

\subsection{Rotationally invariant slave bosons}
\label{sect:methods-RISB}

The RISB approach reformulates the interacting problem in an enlarged Hilbert space containing auxiliary fermions, $\hat{f}$, and slave bosons, $\hat{\Phi}$~\cite{original_risb_2007,risb_impurity_mapping_2015,dmet_risb_dmft_unified_2017}. The physical electrons are mapped to this enlarged space with $\hat{c}^\dagger = \hat{R}[\hat{\Phi}]\hat{f}^\dagger$, where $\hat{R}[\hat{\Phi}]$ is an operator constructed from slave-bosons that encodes the correspondence between local many-body states of the impurity and configurations of the auxiliary fermions~\cite{original_risb_2007}. Lagrange multipliers $\Lambda$ and $E^c$ enforce the local constraints that make this mapping exact, allowing RISB to reproduce the full many-body structure of the original interacting problem. 

A mean-field approximation is required to make the problem computationally tractable. At the mean-field level, the slave bosons condense, $\hat{\Phi} \to \phi$, becoming static complex variational amplitudes. The operator $\hat{R}$ reduces to the renormalization matrix $R$, which rescales and mixes the quasiparticle hopping amplitudes, thereby setting the quasiparticle dispersion and mass enhancement arising from electronic correlations, while the matrix $\Lambda$ serves as a static correlation potential that shifts and mixes the local orbitals.

RISB solutions are obtained self-consistently, with the computational cost dominated by the determination of the ground state of an embedding Hamiltonian at each iteration~\cite{risb_impurity_mapping_2015}. This embedding Hamiltonian describes an impurity coupled to a set of noninteracting bath orbitals~\cite{risb_impurity_mapping_2015}:
\begin{multline}\label{eq:risb_H_emb}
\hat{H}^{\mathrm{emb}} = \hat{H}^{\mathrm{loc}} + \sum_{\alpha}^{N_{\mathrm{p}}} \sum_{a}^{N_{\mathrm{b}}} (D_{a\alpha} \cop^{\dagger}_{\alpha} \fop_a + \textrm{h.c.} ) + \sum_{ab}^{N_{\mathrm{b}}} \Lambda^c_{ab} \fop_a \fop^{\dagger}_b,
\end{multline}
where $\hat{H}^{\mathrm{loc}} = \hat{H}^{\mathrm{int}} + \varepsilon_0$ and $\epsilon_0$ is defined in Eq.~(\ref{eq:epsilon_0}),
$N_{\mathrm{p}}$ is the total number of physical orbitals $\cop$, and $N_{\mathrm{b}}$ is the total number of bath orbitals $\fop$. Here, $D$ is a hybridization coupling the impurity orbitals to the bath, while $\Lambda^c$ is an effective one-body potential acting on the bath orbitals. In RISB, the bath has the same dimension as the impurity ($N_{\mathrm{p}}=N_{\mathrm{b}}$), in contrast to the Anderson impurity model in DMFT, where the bath is formally infinite. This reduced impurity size, combined with only needing to solve for the ground state $\ket{\Phi}$ and evaluate static properties---rather than the fully dynamical Green's function---makes RISB vastly less computationally demanding than DMFT.

The resulting self-energy in RISB has the following mathematical form (see Appendix~\ref{appendix:g-RISB_sigma}):
\begin{equation} \label{eq:self-energy-RISB}
\Sigma(z) = z + \mu - \varepsilon_0 - \bigg[ R^\dagger (z + \mu - \Lambda)^{-1} R \bigg]^{-1}.
\end{equation}

The RISB formalism is known to perform well in the band-insulating and correlated metallic regimes, but to fail to describe correlated insulators. The reason is it retains only the coherent quasiparticle part of the self-energy~\cite{original_risb_2007}, while neglecting higher-energy spectral features such as Hubbard bands. Consequently, RISB accurately captures the Fermi-liquid characteristics of correlated metals---renormalized band dispersions, well-defined quasiparticle peaks, and associated energy shifts---but fails in regimes governed by strong dynamical correlations, such as near the Mott insulator-metal transition, where in RISB the spectral weight vanishes without the formation of Hubbard bands.

\subsection{Ghost rotationally invariant slave bosons (g-RISB)}
\label{sect:methods-g-RISB}

The g-RISB formulation extends RISB by introducing additional ``ghost'' degrees of freedom~\cite{gGA1_2017,original_g-risb}. These extra degrees of freedom enlarge the variational space, enabling g-RISB to capture features beyond the purely coherent quasiparticle description of standard RISB. In practice, g-RISB provides a minimal representation of frequency-dependent self-energy effects, allowing an improved description of strongly correlated regimes~\cite{g-risb_benchmark_1,g-risb_benchmark_2,gGA2_2021,gGA1_2017}, without requiring an explicitly dynamical bath as in DMFT. As a result, g-RISB systematically improves RISB towards DMFT accuracy~\cite{g-risb_benchmark_1,g-risb_benchmark_2}, retaining the computational efficiency of the former---a controllably small bath only requiring the evaluation of a ground state and its static properties---while recovering essential dynamical properties of the latter.

Increasing the number of ghost orbitals enlarges the impurity problem by increasing the total number of bath orbitals $\fop$ entering the embedding Hamiltonian [see Eq.~(\ref{eq:risb_H_emb})], thereby raising the computational cost of each self-consistency iteration. The number of ghost orbitals is controlled by the parameter $\B$, giving $N_{\mathrm{b}}=\B N_{\mathrm{p}}$ auxiliary fermionic degrees of freedom. Hence, $\B$ provides a systematic way to balance accuracy and computational efficiency: larger $\B$ yield a richer variational description with more bath-like hybridization channels through which correlations can be represented, whereas smaller $\B$ captures fewer dynamical correlations, recovering RISB when $\B = 1$. The resulting general algorithmic structure of RISB described in Sec.~\ref{sect:methods-RISB} remains unchanged in g-RISB~\cite{original_g-risb}; only the quasiparticle dimension of the matrices increases, with $R$ and $D$ becoming rectangular.

The self-energy in g-RISB retains the same form as \cref{eq:self-energy-RISB}, but with increased dimension of $R$ and $\Lambda$, resulting in the addition of $N_{\mathrm{g}} = (\B - 1)N_{\mathrm{p}}$ poles to the self-energy (see Appendix~\ref{appendix:g-RISB_sigma_pole_structure}).

Setting $\B = 3$ already remedies several shortcomings of RISB by allowing the self-energy to exhibit two poles. In the correlated metal, the inclusion of two poles allows the theory to capture the position of the Hubbard bands, and simultaneously describe the quasiparticle residue at zero frequency~\cite{gGA1_2017}. In the Mott insulator, one pole collapses to the Fermi level, suppressing low-energy excitations and leaving only the upper and lower Hubbard bands~\cite{gGA1_2017}. Thus, for only a modest increase in computational cost, g-RISB can describe both the metallic and insulating phases, yielding a more accurate picture of the Mott insulator-metal transition~\cite{g-risb_benchmark_1, g-risb_benchmark_2, gGA2_2021}.

\subsection{Numerical details}
\label{sect:methods-num_details}

In this subsection, we provide additional details regarding the calculations that are performed in this work.
The HF and HI calculations were performed using the impurity solvers implemented in the TRIQS 3.3.1 library~\cite{triqs_2015}, the RISB calculations were performed using our own implementation~\cite{RISB_code,Henry_DHL_2021}, and the g-RISB calculations were performed using an implementation part of the TRIQS library which will soon be released. All solutions were converged by requiring the norm difference of successive updates in their free parameters to fall below a specified tolerance: $6\times 10^{-6}$ for HF, $10^{-6}$ for HI, $10^{-10}$ for RISB and $10^{-8}$ for g-RISB. In RISB, the self-energy diverges in the Mott insulator because $R \to 0$, so we added a small diagonal component in $R$.
From the converged self-consistent parameters of each method, we analytically calculated their self-energy $\Sigma$ [see Eqs~(\ref{eq:self-energy-HF}, \ref{eq:self-energy-HI}, \ref{eq:self-energy-RISB})], and along with their converged chemical potentials $\mu$, used these as input to our DMFT calculations [see Eq.~(\ref{eq:g-loc-dmft})].

To solve the impurity at each DMFT iteration, we used continuous-time quantum Monte Carlo (CT-QMC)~\cite{CTQMC_2011}---specifically, the hybridization-expansion variant (CT-HYB)~\cite{werner2006continuous-time,werner2006hybridization}, 
as implemented in the TRIQS 3.3.1 library~\cite{triqs_cthyb_paper, triqs_2015}. They were performed in Matsubara frequencies $i\omega_n$, with Green's functions represented on a 2049 frequency grid, with $n \in [-1024, 1024]$. We chose an intermediate temperature $\beta W = 320$ to ensure that QMC is not too computationally expensive. We took $20$ million QMC measurements with $50000$ warmup cycles on each CPU. The number of cycles between measurements was typically around $20 000$, but was adjusted for each run to ensure the auto-correlation time was approximately unity. The self-energy was post-processed using the default CT-HYB tail-fitting parameters: the high-frequency moments were calculated from the impurity density matrix as in~\cite{asymptoticSigmaWang2011, stable_Sigma_calc_2025}, and the last 20\% of the self-energy was replaced with a fitted tail.

We evaluated lattice integrals either through Brillouin-zone integration or through energy-space integration, depending on the method. In HF and RISB, a $100 \times 100$ $k$-point grid and an inverse temperature $\beta W = 320$ were used for Brillouin-zone integration. In HI and DMFT, integrals of lattice Green’s functions were transformed to integrals over energy via a Hilbert transform, constructed from a discretized density of states on a $1000$-point energy grid, itself generated by integrating over a $10\,000 \times 10\,000$ $k$-point grid of the first Brillouin zone.

Real-frequency spectra were obtained by maximum entropy analytic continuation~\cite{maxentJarrell1996} using the version 1.2 implementation in the TRIQS library~\cite{maxent}. The Bryan cost function was used, the optimal value of the hyperparameter $\alpha$ was found with the line-fit analyzer, and the Green's functions were continued to a hyperbolic real-frequency mesh with $200$ points for $\omega/W \in [-3.125, 3.125]$.

\section{Results}
\label{sect:results}

Having introduced the methods, we are now ready to compare them. 
In Sec.~\ref{sect:conv_comp_mu}, we assess the performance of our approach in the strongly correlated regime ($U / W = 1.875$) for fixed chemical potentials in four representative regimes: the hole-doped metal ($\mu/W = 0.375$), the Mott transition ($\mu/W = 0.5$), the Mott insulator ($\mu/W = 0.875$), and the electron-doped metal ($\mu/W = 1.5$). In particular, we compare this starting self-energy and impurity Green's function with the converged one from DMFT, and then we study the DMFT convergence as a function of the starting point.
In Sec.~\ref{sect:conv_comp_nfix}, we perform a similar comparison, but now fixing the electronic filling. We study the half-filling case just below the Mott transition ($n=1, U/W=1$) and a strongly correlated hole-doped metal ($n=0.85, U/W=1.5$).
Finally in Sec.~\ref{sect:one-shot}, we compare the spectral functions obtained by initializing DMFT with the various quantum embedding methods and performing a single shot DMFT calculation. Our results suggest that, for a sufficiently accurate initial solution like the one provided by g-RISB with $\B=3$, only a single-shot DMFT is required to obtain a reasonable spectral function.

\subsection{Convergence comparison at fixed chemical potential}
\label{sect:conv_comp_mu}

\begin{figure*}
    \centering
    \includegraphics[width=\linewidth]{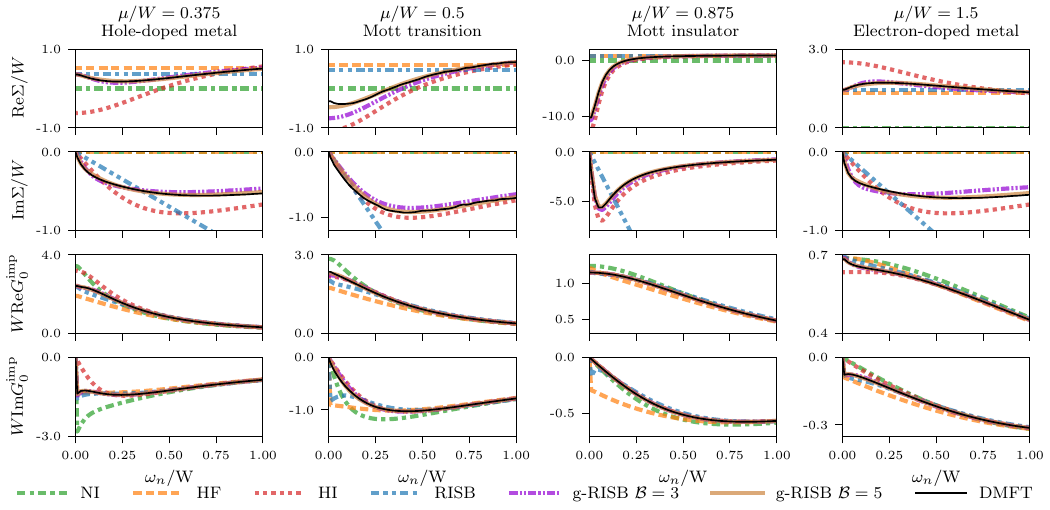}
    \caption{\label{fig:weiss_sigma_fixed_mu}
    Comparison of the self-energies and Weiss fields obtained from converged solutions.
    When the low-frequency part of the self-energy $\Sigma$ (rows 1 and 2) is accurate, the impurity model defined through the Weiss field $G_0^{\mathrm{imp}}$ (rows 3 and 4) is very close to the converged DMFT one. In contrast, high-frequency errors in $\Sigma$ largely cancel in $G_0^{\mathrm{imp}}$ (cf. Eq.~(\ref{eq:sigma_cancellation})), as illustrated by the performance of RISB in the metals. Results shown are at fixed chemical potential $\mu$ for the paramagnetic square-lattice Hubbard model ($t'/t = 0.15$ and $U/W=1.875$); the solid black line denotes converged DMFT, and $W$ is the non-interacting bandwidth.
    }
\end{figure*}

When performing a DMFT calculation, the object that defines the impurity problem is the so-called Weiss field. As discussed in Sec.~\ref{sect:methods}, the Weiss field for a fixed dispersion $\epsilon_{\textbf{k}}$ and chemical potential is only tuned by on the provided self-energy. In Fig.~\ref{fig:weiss_sigma_fixed_mu}, we present the converged self-energies $\Sigma(i\omega_n)$ and resulting Weiss fields $G^{\text{imp}}_0(i\omega_n)$ for each of the method considered, with the addition of g-RISB with $\B=5$, and compare them with the converged DMFT objects. We do so in four different regimes: the hole-doped metal, the Mott transition, the Mott insulator and the electron-doped metal.

Overall, we find that, as anticipated from Eq.~(\ref{eq:sigma_cancellation}), even when the high-energy part of the self-energy is inaccurate, the low-energy part suffices to yield an accurate Weiss field. More specifically, the HF and RISB methods reproduce the Weiss field more accurately than the NI one in the metallic regimes, with RISB outperforming HF. However, neither of them offers a significant improvement in the insulator regime. The HI method, by contrast, performs well in insulators but is comparable to NI in metals. The method that shines in both regimes is g-RISB, which provides the most accurate self-energies and Weiss fields overall. It is only near the Mott transition, as for all methods, that it shows a reduced ability to match the Weiss field. Note however that that increasing $\B=5$ in g-RISB improves those results, showing that this methods can be systematically improved by increasing $\B$.

We now examine how improving the Weiss-field initialization influences the convergence of DMFT by examining its iteration-by-iteration behavior. Our analysis focuses on physically relevant quantities that characterize the Mott transition: the kinetic energy per site
\begin{align}
    E_{\text{kin}} & = \frac{1}{N} \sum_{\bm{k}\sigma n} \varepsilon_{\bm{k}} \braket{\nop{\bm{k}\sigma}} \nonumber \\
    & = \frac{1}{N \beta} \sum_{\bm{k}\sigma n} \varepsilon_{\bm{k}} \frac{1}{i\omega_n + \mu - \varepsilon_{\bm{k}} - \Sigma(i\omega_n)} e^{i\omega_n 0^+},
\end{align}
the potential energy per site
\begin{equation}
    E_{\text{pot}} = \frac{U}{N} \sum_i \braket{\nop{i\uparrow} \nop{i\downarrow}} = U \braket{\nop{\uparrow} \nop{\downarrow}}_{\mathrm{imp}},
\end{equation} 
the impurity filling
\begin{equation}
    n_{\text{imp}} = \sum_{\sigma} \braket{\nop{\sigma}}_{\mathrm{imp}},
\end{equation}
and the Matsubara-frequency approximation for the quasiparticle weight~\cite{QP_weight_from_Matsubara}
\begin{align}
        Z & = \bigg[ 1 - \frac{\partial \Re \Sigma(\omega)}{\partial \omega} \bigg|_{\omega=0} \bigg]^{-1}
        \approx \bigg[ 1 - \frac{\Im \Sigma(i\omega_0)}{\omega_0} \bigg]^{-1}.
\end{align}
Above, $\braket{\cdot}_{\mathrm{imp}}$ denotes an expectation value evaluated from the impurity within CT-QMC, and the infinitesimal factor $e^{i\omega_n 0^+}$ ensures convergence of Matsubara sums.

To capture convergence in spectral properties we use the Frobenius norm difference of the Weiss field between successive iterations, defined as
\begin{equation}\label{eq:convergence_definition}
\mathcal{C}_\ell = \sqrt{\sum_n \mathrm{Tr} [\delta_{\ell}(i\omega_n)^\dagger \delta_{\ell}(i\omega_n)]},
\end{equation}
where $\ell$ is the iteration number and $\delta_\ell(i\omega_n) = G_0^{\mathrm{imp,\, \ell}}(i\omega_n) - G_0^{\mathrm{imp,\,\ell-1}}(i\omega_n)$. 

\begin{figure*}
	\begin{centering}
		\includegraphics[width=\linewidth]{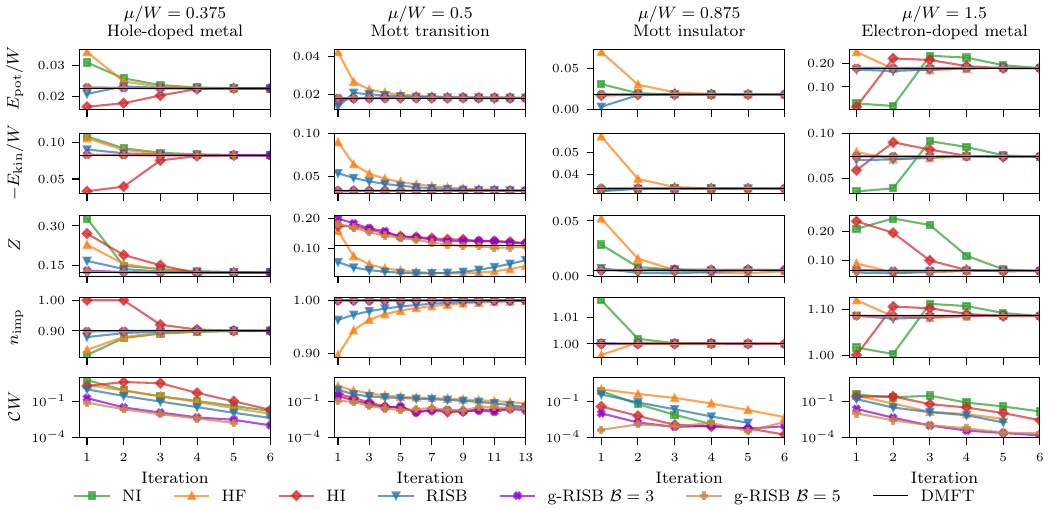}
		\caption{\label{fig:obs_fix_mu}
        DMFT convergence as a function of its initialization in four different regimes of fixed chemical potential.
        A better initialization markedly improves convergence across metallic and insulating regimes, as well as near the Mott phase transition. Among them, g-RISB performs best because it reproduces the low-energy correlation effects with accuracy comparable to DMFT, cf. Fig.~\ref{fig:weiss_sigma_fixed_mu}. Here we plot the kinetic energy, $E_{\mathrm{kin}}$, potential energy, $E_{\mathrm{pot}}$, quasiparticle weight, $Z$, impurity filling, $n_{\mathrm{imp}}$, and norm difference of the Weiss field between successive iterations, $\mathcal{C}$.
		}
	\end{centering}
\end{figure*}

Figure~\ref{fig:obs_fix_mu} shows the convergence of DMFT in the four different regimes. The key observation is straightforward: a better initialization not only reduces the number of iterations by starting DMFT near its converged values, but also yields a more robust self-consistent procedure than NI-based initialization. Indeed, NI behaves inconsistently across regimes: typically it begins farthest from the fixed point, it performs especially poorly in the electron-doped metal, and it fails entirely at the Mott transition due to the sign problem.

A fast convergence in the metallic regimes requires an initial Weiss field that captures the quasiparticle renormalization, and it is further improved by describing the accompanying Hubbard bands. g-RISB satisfies both criteria and converges the fastest; RISB (renormalization but no the Hubbard bands) and HF lag behind; HI and NI perform the worst. This ranking is especially pronounced on the electron‑doped side, where NI erroneously predicts an insulator and therefore starts farthest from the fixed point.

Inside the Mott insulator, an atomic‑like structure for $\Sigma$ that places the Hubbard bands at the correct energies is already near the DMFT fixed point. Consequently, HI and g-RISB converge almost immediately, with the former being particularly appealing on cost grounds when the study is restricted to the insulating regime.

At the Mott transition, initializing DMFT with a correlated method enables convergence even when starting far from the fixed point, while the non-interacting initialization fails to converge due to a severe sign problem. In this regime, convergence is ultimately limited by QMC noise: the Weiss-field norm, $\mathcal{C}$, plateaus around $10^{-2}$, reflecting a numerical noise floor arising from fluctuations in spectral features between successive iterations. This residual noise slows convergence and constrains accuracy. The most efficient convergence is achieved when the initialization begins close enough to the DMFT fixed point to be contained within the noise.

The increased computational cost at the Mott transition reflects both the generally slower convergence near a phase transition~\cite{RozenbergCriticalSlowingDMFT1999} and the inability of the initial solutions to accurately capture the relevant low-energy features, as shown in Fig.~\ref{fig:weiss_sigma_fixed_mu}. Even g-RISB, the best initial solution, reproduces the self-energy accurately only above $W/4$, but it deviates at the lowest Matsubara frequencies that determine the quasiparticle renormalization. These modest low-energy discrepancies are sufficient to substantially slow convergence near the Mott transition (cf. the convergence of $Z$ and $\mathcal{C}W$ in Fig.~\ref{fig:obs_fix_mu}).

The sensitivity to low-energy accuracy also explains why improving the quality of g-RISB initialization systematically enhances convergence. Proximity to the DMFT fixed point is tunable within g-RISB, where increasing the number of ghost orbitals $\mathcal{B}$ brings the starting point progressively closer to convergence. With $\mathcal{B}=3$, g-RISB reproduces converged DMFT results in all regimes after a single iteration, except at the Mott transition where $Z$ requires additional steps. Increasing to $\mathcal{B}=5$ brings the initialization even closer to the fixed point, as reflected in a smaller $\mathcal{C}W$ at the first iteration and improved performance at the Mott transition.

\begin{figure}
	\begin{centering}
		\includegraphics[width=\linewidth]{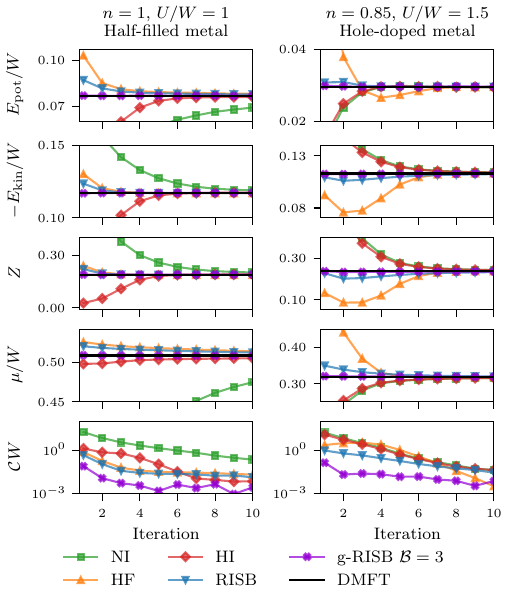}
		\caption{\label{fig:obs_fixed_n}
        DMFT convergence as a function of its initialization at two different electronic fillings. Targeting a fixed electron filling $n$ makes to choice convergence more sensitive to the choice of initial solutions, since enforcing charge conservation introduces an additional nonlinear feedback between the chemical potential and the self-energy. The initialization with g-RISB starts DMFT very near its converged solution, so that remaining iterations only adjust minor spectral features (as $\mathcal{C}W$ continues to decrease), keeping the total cost comparable to the fixed–chemical-potential case. Results are shown for metals at half filling ($n=1$, $U/W=1$) and hole-doped from half filling ($n=0.85$, $U/W=1.5$).
		}
	\end{centering}
\end{figure}

\subsection{Convergence comparison at fixed orbital filling}
\label{sect:conv_comp_nfix}

Having examined convergence at fixed chemical potential, we now study it at fixed electron filling $n$, as many applications require constraining the filling according to the chemistry of the material. The target filling is enforced by adjusting $\mu$ at each DMFT iteration so that the local Green's function $G^{\mathrm{loc}}$ from Eq.~(\ref{eq:g-loc-dmft}) yields the desired $n$. 

Fixed filling adds an additional nonlinear constraint to the self-consistency loop, so the quantum embedding solutions used to initialize the DMFT loop must provide not only a reasonable self-energy $\Sigma$, but also a good initial chemical potential $\mu$. Because $\mu$ must be re-tuned at every iteration, the feedback between $\Sigma$ and $\mu$ introduces further strong nonlinearity into the DMFT self-consistency loop, which can slow convergence and increases its sensitivity to the quality of the initialization.

The benefit of an accurate initial solution becomes even more pronounced at fixed filling compared to the fixed–chemical-potential case. In Fig.~\ref{fig:obs_fixed_n}, we show two metallic regimes: one at half filling just below the Mott transition ($n=1$, $U/W=1$) and another hole-doped one with strong interactions ($n=0.85$, $U/W=1.5$). We clearly see that initializing with NI now requires substantially more iterations than with the other quantum embedding methods, especially at half filling, because it starts DMFT far from convergence. The increased advantage of RISB over HF, particularly in the hole-doped metal, emphasizes the importance of an initial solution that captures the quasiparticle renormalization in metallic regimes. Across all correlated regimes, g-RISB remains the most accurate and effective method to precondition DMFT.

We find that g-RISB can yield an order-of-magnitude reduction in the computational cost required to reach converged DMFT compared to other initial solutions. This efficiency arises because g-RISB places DMFT extremely close to its converged solution, so that only minor refinements are still required and the calculation is effectively converged after a single iteration. The advantage is especially pronounced at half filling, where $\mu$ requires more than ten iterations to converge for the other initial solutions.

Overall, the performance of the quantum embedding methods has a consistent ranking across correlated regimes. g-RISB is the most effective, typically placing DMFT close to its fixed point after a single iteration; RISB performs well in metallic regimes and consistently outperforms HF; HI excels in insulating phases but performs poorly in metals; and NI is the least efficient and reliable.

\subsection{One-shot DMFT}
\label{sect:one-shot}

\begin{figure*}
	\begin{centering}
		\includegraphics[width=\linewidth]{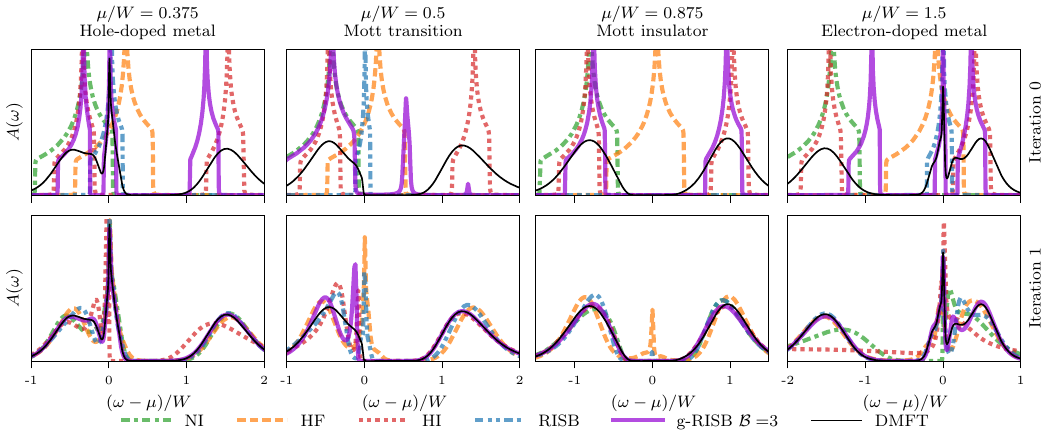}
		\caption{\label{fig:spectral_it1_fixed_mu}
Demonstration that a single DMFT iteration initialized with a computationally cheaper method can closely reproduce the converged DMFT spectral function. Among the tested approaches, g-RISB gives the closest agreement across all correlated regimes. Near the Mott transition all methods deviate from the converged DMFT spectra because of small low-frequency errors in the self-energy that affect the Weiss field (cf. Fig.~\ref{fig:weiss_sigma_fixed_mu}). The spectral functions $A(\omega)$ at fixed chemical potential $\mu$ are shown before (top: Iteration 0) and after one DMFT step (bottom: Iteration 1), obtained via maximum-entropy analytic continuation.
	}
	\end{centering}
\end{figure*}

The rapid convergence of DMFT when initializing it with a quantum embedding solutions from another method suggests that a single DMFT iteration can already recover most of the dynamical correlations absent in the cheaper solution. Because the Weiss field depends primarily on the low-frequency part of the self-energy (see~Eq.~(\ref{eq:sigma_cancellation})), an initial solution that accurately captures the low-energy physics places the system close to the DMFT fixed point. A single DMFT iteration then corrects the remaining deficiencies through feedback between the impurity and its dynamical environment---restoring the high-frequency asymptotic behavior of the self-energy, adding incoherent spectral weight, and providing access to dynamical observables---while avoiding the cost of full self-consistency.

To assess the effectiveness of one-step DMFT, we compared the local spectral function after one DMFT iteration with the converged one. Figure~\ref{fig:spectral_it1_fixed_mu} shows the spectra for the four cases considered at fixed $\mu$ in Sec.~\ref{sect:conv_comp_mu}, both before (top panels) and after (bottom panels) the DMFT iteration. The converged spectra exhibit the characteristic signatures of strong correlations: well-defined incoherent lower and upper Hubbard bands and, in metallic phases, a renormalized quasiparticle peak at the Fermi level---features that one-step DMFT must reproduce. Before any DMFT iterations, none of the initial spectra contains the full dynamical structure obtained in the converged solution.

In the metallic regimes, the success of one-step DMFT hinges on how accurately the initial solution captures the quasiparticle peak. A single DMFT iteration, initialized with RISB or g-RISB---both of which capture quasiparticle renormalization---yields spectral functions that are qualitatively indistinguishable from fully self-consistent DMFT on both the hole- and electron-doped sides. By contrast, HF and NI overestimate the quasiparticle weight and miss fine spectral features, particularly in the electron-doped regime. HI performs poorly in both metals: although its standalone solution places the Hubbard bands at roughly the correct energies, it starts so far from the DMFT fixed point that a single iteration is insufficient to reach the convergence basin, causing the Hubbard bands to vanish on the electron-doped side and the quasiparticle peak to remain absent on the hole-doped side.

In the Mott-insulating regime, the situation is reversed: the accuracy of the positions of the Hubbard bands governs the success of one-step DMFT. In this case, HI and g-RISB perform exceptionally well, because both capture the atomic multiplet structure and recover the correct insulating gap after a single DMFT iteration. By contrast, NI, HF, and RISB, which fail to describe the Hubbard bands, produce metallic or weakly gapped spectra and do not reproduce the insulating behavior.

At the Mott transition, the performance of one-step DMFT is reduced because the system lies at the boundary between coherent and incoherent regimes, where small inaccuracies in the low-energy part of the self-energy can strongly affect its proximity to the DMFT fixed point. One-step DMFT initialized with HF or RISB incorrectly retains a residual quasiparticle peak, whereas initialization with HI or g-RISB removes the peak but places excess spectral weight just below the Fermi level. Achieving quantitative agreement at the transition typically requires additional DMFT iterations (see Figs~\ref{fig:obs_fix_mu} and \ref{fig:obs_fixed_n}) or, in the case of g-RISB, more ghost orbitals.

\begin{figure}
	\begin{centering}
		\includegraphics[width=\linewidth]{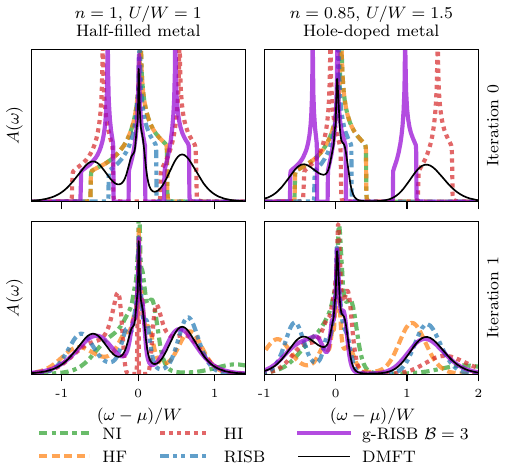}
		\caption{\label{fig:spectral_fixed_n}
        Comparison of the spectra before and after the DMFT iteration for the fixed filling cases. In these cases, one-step DMFT only accurately reproduces the full DMFT spectrum when initialized with a correlated method that captures the essential low-energy physics. Cheaper initial methods such as NI, HF, and HI fail to recover the correct quasiparticle and Hubbard-band structure, while RISB is qualitatively correct but quantitatively poor. Initialization with g-RISB yields spectra nearly indistinguishable from the converged DMFT.
		}
	\end{centering}
\end{figure}

Results for the cases at fixed $n$ considered in Sec.\ref{sect:conv_comp_nfix}, shown in Fig.~\ref{fig:spectral_fixed_n}, exhibit the same behavior as one-step DMFT at fixed $\mu$, but with sharper contrasts between the different initial solutions. When the initial $\Sigma$ and $\mu$ are already accurate, a single DMFT iteration suffices to recover the full spectral function. NI and HF fail to produce reliable spectra because their starting chemical potentials deviate significantly from the DMFT value, leading to misplaced bands, incorrect gaps, and quasiparticle peaks with incorrect weight and width. RISB, which captures only the quasiparticle peak, restores the Hubbard bands after one iteration but remains quantitatively inaccurate. HI places the Hubbard bands roughly correctly but provides a poor chemical potential and lacks coherent spectral weight near the Fermi level. One-step DMFT initialized with g-RISB reproduces both the quasiparticle and Hubbard-band structures with near-quantitative accuracy. 

Our results establish a clear guiding principle: when the initial self-energy and chemical potential place the system near the DMFT fixed point, a single DMFT iteration is sufficient to reconstruct the missing dynamical structure and achieve near–DMFT accuracy. In contrast, poor initial estimates of the low-energy physics produce spectra that remain qualitatively incorrect after one iteration. However, even if the initial solution omits key spectral components---such as the Hubbard bands absent in RISB solutions---a single DMFT step often restores them, yielding spectra that are qualitatively correct and, in many cases, close to the fully converged result.

\section{Conclusions}
\label{sect:concl}

In this work, we have shown that using computationally cheaper methods to initialize the DMFT self-consistency cycle can substantially accelerate convergence and greatly reduce computational cost, with the benefit becoming even more pronounced at fixed electron filling. The more the initial solutions captures correlated low-energy effects on the scale of the non-interacting bandwidth, the more the DMFT convergence is improved. In the best case---realized with g-RISB and providing an order-of-magnitude reduction in computational cost---a single DMFT iteration produces observables and spectra that are indistinguishable from those of fully converged DMFT.

We expect that the benefits of our scheme will also transfer to larger systems, and therefore will have an even greater impact, since the computational cost of DMFT scales more steeply with system size. By delegating the task of bringing the DMFT self-consistency cycle close to its fixed point to a cheaper solution, the resulting reduction in overall computational cost can be substantially larger than in small systems. This applies to a range of problems, such as multi-orbital systems~\cite{DMFT_book}, multiple decoupled impurities~\cite{decoupled_impurities_1999, Jaime_decoupled_dmft}, and to cluster extensions~\cite{DMFT_book}, which we leave for future works.

Another particularly promising application of our approach is \textit{ab initio} density functional theory (DFT) combined with DMFT (DFT+DMFT), where computational costs are especially high. In this context, a DFT+DMFT calculation can be initialized from a converged, computationally less expensive, charge–self-consistent quantum embedding method~\cite{CuiDFT+DMET2020,dft+risb_original,dft+g-risb_original}. Because, for example, the charge density obtained from charge–self-consistent DFT+g-RISB calculations closely matches that of full DFT+DMFT~\cite{dft+g-risb_original}, performing a single DMFT iteration on top of this initialization offers a promising route to accessing dynamical correlations at significantly reduced computational cost, while maintaining accuracy comparable to fully self-consistent DFT+DMFT.

Moreover, we expect that initializing DMFT with another quantum embedding method may help mitigate the sign problem in QMC studies, which can severely limit the practical applicability of DMFT. Initializing DMFT far from its fixed point forces the self-consistency cycle to traverse a broad region of the parameter space, thereby increasing the likelihood of encountering regimes with a severe sign problem. This sensitivity arises because, in QMC, the average sign depends on the free-energy difference $\Delta F$---scaling as $\exp(-\beta \Delta F)$---between the system and an auxiliary bosonic system introduced to enforce positive Monte Carlo weights~\cite{CTQMC_2011}. The weights depend on the Weiss field~\cite{CTQMC_2011}, so that certain values encountered during the self-consistent loop can lead to a much more severe sign problem than others. Provided the sign problem remains manageable, a better initialization can reduce this risk by starting DMFT near its fixed point and avoiding regions of parameter space where the sign problem is severe.

Overall, our results reinforce a clear strategy for reducing the computational cost in iterative methods for many-body problems: initialize the self-consistency cycle with computationally cheaper approximations that place the system close to its fixed point. In the context of DMFT, this approach yields substantial reductions in computational cost and, in particular, opens a practical route toward accurate high-throughput screening of strongly correlated materials for the design of future quantum technologies.

\section*{Acknowledgments}

We thank Alexander Hampel and Jaime Merino for helpful conversations. EM was supported by an Australian Government Research Training Program (RTP) Scholarship and a Queensland Government Department of Environment, Science and Innovation top-up scholarship. This work was supported by the Australian Research Council (DP230100139, FT230100653). The Flatiron Institute is a division of the Simons Foundation. N.L. gratefully acknowledges funding from the National Science Foundation under Award No. DMR-2532771 and from the Simons Foundation (Grant No. 00010910). T.-H.L. gratefully acknowledges funding from the National Science and Technology Council (NSTC) of Taiwan under Grant No. NSTC 112-2112-M-194-007-MY3 and the National Center for Theoretical Sciences (NCTS) in Taiwan.

\section*{Appendices}

\appendix
\section{g-RISB self-energy}
\label[appendix]{appendix:g-RISB_sigma}

The Green's function in g-RISB is 
\begin{equation}\label{eq:g-RISB_Greens_function}
G(z) = \frac{1}{z - \varepsilon_k - \Sigma(z)}
= R^\dagger \frac{1}{z - H_{\mathrm{qp}}} R,
\end{equation}
where $H_{\mathrm{qp}} = R\varepsilon_kR^\dagger - \Lambda$ is the quasiparticle Hamiltonian. Our goal is to take the inverse of the RHS to obtain an analytic form for the self-energy $\Sigma(z)$. Because $R$ is rectangular, a straightforward inversion is not possible. Instead, we will make use of the following lemma:

\begin{lemma}[Woodbury inversion identity]\label{lemma:woodbury}
    For $n\times n$ matrix $A$, $m\times m$ $C$, $n\times m$ $U$ and $m\times n$ $V$,
    \begin{multline}
        (A \pm UCV)^{-1} = A^{-1}
        \\
        \mp A^{-1}U(C^{-1} \pm VA^{-1}U)^{-1}VA^{-1}.
    \end{multline}
\end{lemma}

To invert the Green's function, we first apply the Woodbury lemma with $A = z - \Lambda$, so that
\begin{multline}
    ( A - R\varepsilon_k R^\dagger )^{-1} = A^{-1} 
    \\
    + A^{-1} R (\varepsilon_k^{-1} - R^\dagger A^{-1} R )^{-1} R^{\dagger}A^{-1}.
\end{multline}
Multiplying from the left and right by $R^\dagger$ and $R$, respectively, and defining $B = R^\dagger A^{-1} R$, we have
\begin{align}
    \nonumber G(z) &= B + B (\varepsilon_k^{-1} - B)^{-1} B
    \\
    \nonumber &= B \big[ 1 + (\varepsilon_k^{-1} - B)^{-1} B \big]
    \\
    \nonumber &= B (\varepsilon_k^{-1} - B)^{-1} \big[ (\varepsilon_k^{-1} - B) + B \big]
    \\
    &= B (\varepsilon_k^{-1} - B)^{-1} \varepsilon_k^{-1}.
\end{align}
This form can be easily inverted, yielding
\begin{equation}
    G(z)^{-1} = \varepsilon_k (\varepsilon_k^{-1} - B) B^{-1} = B^{-1} - \varepsilon_k.
\end{equation}
Replacing the original variables, we have
\begin{equation}
    G(z)^{-1} = \big[ R^\dagger (z - \Lambda)^{-1} R \big]^{-1} - \varepsilon_k.
\end{equation}
Comparing to~Eq.~(\ref{eq:g-RISB_Greens_function}), the self-energy is
\begin{equation}\label{eq:sigma_from_woodbury}
    \Sigma(z) = z - \big[ R^\dagger (z - \Lambda)^{-1} R \big]^{-1},
\end{equation}
which is the form presented in the main text.

\section{Pole structure of the g-RISB self-energy}
\label[appendix]{appendix:g-RISB_sigma_pole_structure}

To gain further insight into the pole structure of the self-energy, we define $F^2 = \Lambda$ and apply the Woodbury inversion lemma several times to~Eq.~\ref{eq:sigma_from_woodbury}. Below, we indicate the terms being used in the Woorbury inversion lemma with curly braces for clarity. With this notation,
\begin{multline}
    (z - \Lambda)^{-1} = (\{z\} - \{F\}\{1\}\{F\})^{-1}
    \\
    = z^{-1} + z^{-1} F (1 - F z^{-1} F)^{-1} F z^{-1}.
\end{multline}
Defining $S = R^\dagger R$, the self-energy is
\begin{multline}
    \Sigma(z) = z - \big[ \{z^{-1}S\} 
    \\
    + \{z^{-1} R^\dagger F\} \{(1 - F z^{-1} F)^{-1}\} \{F R z^{-1}\} \big]^{-1}.
\end{multline}
Applying Woodbury again gives
\begin{multline}
    \Sigma(z) = (1 - S^{-1})z - S^{-1} R^\dagger F \times
    \\
    (1 - z^{-1}FF + z^{-1} FRS^{-1}RF)^{-1} F R S^{-1},
\end{multline}
which may be written as 
\begin{multline}
    \Sigma(z) = (1 - S^{-1})z - S^{-1} R^\dagger F \times
    \\
    (1 - z^{-1} F [1 - R S^{-1} R^\dagger] F )^{-1} F R S^{-1}.
\end{multline}
We define the projector $Q = 1 - RS^{-1}R^\dagger = Q^2$ and re-write the self-energy as
\begin{multline}
    \Sigma(z) = (1 - S^{-1})z - S^{-1} R^\dagger F \times
    \\
    (\{1\} - \{FQ\}\{z^{-1}\}\{QF\})^{-1} F R S^{-1}.
\end{multline}
A final application of the Woodbury inversion lemma gives
\begin{multline}
    \Sigma(z) = (1 - S^{-1})z - S^{-1} R^\dagger F \times
    \\
    (1 + FQ(z - QFFQ)^{-1}QF ) F R S^{-1}.
\end{multline}
Recalling that $F^2 = \Lambda$, the self-energy may be written
\begin{multline}\label{eq:sigma_pole_structure}
    \Sigma(z) = (1 - S^{-1})z - S^{-1}R^\dagger \Lambda RS^{-1} 
    \\
    + S^{-1}R^\dagger \Lambda Q \frac{1}{z - Q\Lambda Q} Q\Lambda R S^{-1},
\end{multline}
which shows that the poles of the self-energy come from the subspace projected to by $Q$.


The projector $Q$ is onto the `ghost subspace' of dimension $N_{\mathrm{g}} = N_{\mathrm{b}} - N_{\mathrm{p}}$. To show this, we identify $R^+ = (R^\dagger R)^{-1} R^{\dagger}$ as the unique pseudoinverse (Moore-Penrose inverse) of $R$. By the definition of the pseudoinverse, $RR^+$ projects onto the image of $R$: the span of the columns of $R$. Under the assumption that $R$ has full column rank, this space has the same dimension as the number of physical electron orbitals, $N_{\mathrm{p}}$, and we identify it with the embedding of the physical subspace in the full quasiparticle space. The projector onto the orthogonal complement---the ghost subspace---is then $Q = 1 - RR^+$.

Due to projecting with $Q$, the pole term $Q(z - Q\Lambda Q)^{-1}Q$ in Eq.~\ref{eq:sigma_pole_structure} allows at most $N_{\mathrm{g}}$ poles in the self-energy. This number is maximized when $Q \Lambda Q$ has no repeated eigenvalues except the $N_p$ zero eigenvalues coming from the projection, while the $N_{\mathrm{p}}$ divergences induced by these zero eigenvalues are projected out by the outer $Q$.

This is a gauge-invariant generalization of the fixed-gauge pole expansion of previous work~\cite{g-risb_benchmark_2}. To show this, we introduce a gauge transform $u$ such that 
\begin{equation}
    u^\dagger R = \tilde{R} = \begin{pmatrix}
        \tilde{R}_0 \\
        0
    \end{pmatrix}, \qquad u^\dagger \Lambda u = \tilde{\Lambda} = \begin{pmatrix}
        \tilde{\lambda}_0 & \tilde{\lambda}_1 \\
        \tilde{\lambda}_1^\dagger & \tilde{\lambda}_2
    \end{pmatrix},
\end{equation}
as in Ref.~\cite{g-risb_benchmark_2}, where $\tilde{R}_0$ and $\tilde{\lambda}_0$ are $N_{\mathrm{p}} \times N_{\mathrm{p}}$ matrices, $\tilde{\lambda}_1$ is $N_{\mathrm{p}} \times N_{\mathrm{g}}$, and $\tilde{\lambda}_2$ is $N_{\mathrm{g}} \times N_{\mathrm{g}}$. A straightforward evaluation shows that in this gauge, the projector $Q$ is in its eigenbasis and is given by 
\begin{equation}
    \tilde{Q} = \begin{pmatrix}
        0 & 0 \\
        0 & 1
    \end{pmatrix},
\end{equation}
where the identity block has size $N_{\mathrm{g}} \times N_{\mathrm{g}}$, corresponding to the $N_{\mathrm{g}}$ eigenvalues of unity of $Q$. Thus the self-energy in this gauge is
\begin{multline}
    \tilde{\Sigma}(z) = z (1 - [\tilde{R}_0^\dagger \tilde{R}_0]^{-1}) + \tilde{R}_0^{-1} \tilde{\lambda}_0 [\tilde{R}_0^\dagger]^{-1} 
    \\
    + \tilde{R}_0^{-1} \tilde{\lambda}_1 \frac{1}{z - \tilde{\lambda}_2} \tilde{\lambda}_1^\dagger [\tilde{R}_0^\dagger]^{-1},
\end{multline}
which is the same form derived in Eq. 17 of Ref.~\cite{g-risb_benchmark_2}.

\bibliography{main.bib}

\end{document}